\documentstyle[titlepage,twoside,12pt]{article}
\begin{document}

\pagenumbering{arabic}
\setcounter{page}{1}

\pagenumbering{arabic}

{\LARGE \bf A question in the axiomatic approach} \\
{\LARGE \bf to Quantum Mechanics} \\ \\

{\bf Elem\'{e}r E Rosinger \\ Department of Mathematics \\ University of Pretoria \\
Pretoria, 0002 South Africa \\
e-mail : eerosinger@hotmail.com} \\ \\

{\bf Abstract} \\

The classical Hilbert space formulation of the axioms of Quantum Mechanics appears to leave
open the question whether the Hermitian operators which are associated with the observables of
a finite non-relativistic quantum system are {\it uniquely} determined. \\ \\

{\bf 1. Introduction} \\

For simplicity, we shall only consider the basic axiomatic formulation of the non-relativistic
Quantum Mechanics of a finite systems in which the operators $A$ associated with the
observables \\

(1) $ \begin{array}{l} ~~~\mbox{are everywhere defined bounded Hermitian operators
                                  on a given} \\
                       ~~~\mbox{Hilbert space $H$ which corresponds to the quantum system,}
     \end{array} $ \\ \\

(2) $ \begin{array}{l} ~~~\mbox{have a countable orthonormal complete set of eigenvectors} \\
                       ~~~ \alpha_n \in H,~ \mbox{with}~~ n \in {\bf N},
      \end{array} $ \\ \\

(3) $ \begin{array}{l}  ~~~\mbox{with the corresponding eigenvalues}~ A_n \in {\bf R},
                                    ~\mbox{with}~~ n \in {\bf N}, \\
                        ~~~\mbox{pair-wise different}.
      \end{array} $ \\ \\

Such a basic framework, Gillespie, is still capable to point to the question of our concern.
Here we recall the relevant axioms on Measurement, Gillespie [pp. 49-58]. \\
Given a moment of time $t \in {\bf R}$, let us assume that immediately prior to that moment,
the quantum system has been in the state \\

(4) $~~~~~~ \psi_{t - 0} \in H $ \\

Further, let us assume given a certain physical observable ${\cal A}$ with the corresponding
Hermitian operator \\

(5) $~~~~~~ A \in {\cal L} ( A ) $ \\

under assumtions (1) - (3), where ${\cal L} ( A )$ denotes the set of all bounded linear
operators on $H$. \\
Then according to one of the axioms, Gillespie [p. 49], the strongest predictive statement we
can make about the result of the measurement following the observation ${\cal A}$ of the
system effectuated at time $t$ is given by the sequence of relations \\

(6) $~~~~~~ \mbox{Prob} ( A_n ) ~=~ | < \alpha_n, \psi_{t - 0} > |^2,~~~ n \in {\bf N} $ \\

Another axiom, Gillespie [p. 58], states that immediately after the moment $t$ when the
observation ${\cal A}$ was effectuated on the system, the state of the system will collapse to
the state \\

(7) $~~~~~~ \psi_{t + 0} ~=~ \alpha_n $ \\

where $n \in {\bf N}$ is precisely the index of the eigenvalue $A_n$ which happened to be
realized by the given measurement following the observation ${\cal A}$ was effectuated on the
system at time $t$. And in view of the assumption (3), this index $n$ is well
defined. \\ \\

{\bf 2. A first formulation of the question} \\

We recapitulate. Given a physical observable ${\cal A}$, then the
association \\

(8) $~~~~~~ {\cal A} ~\longmapsto~ A \in {\cal L} ( H ) $ \\

under the assumptions (1) - (3), is only required to satisfy (6) and (7). \\

We note that the operator $A$ is uniquely determined by its eigenvectors and eigenvalues in
(2) and (3), respectively. Further, in view of the collapse in (7), it follows that the
eigenvectors in (2) are uniquely determined by the observable ${\cal A}$. \\

Here it is important to note that, far as the eigenvalues (3) are concerned, the {\it only}
condition they are supposed to satisfy is (6). However, due to the presence of {\it
probabilities} in (6), it is {\it not} certain that the eigenvalues (3) will also be uniquely
determined by that relation. \\

Let us, therefore, consider relation (6) more carefully about its possible precise meaning. \\

First, (6) does not refer directly to the actual value of any of the eigenvalues $A_n$, but
only to the probability of the occurrence in measurement. Second, the relation (6) cannot in
general be perfectly be confirmed or verified in effective practical experiments, since that
would involve the performance of very large numbers, if not in fact, of infinitely many
measurements. Third, an effective practical measurement cannot in general give the exact value
of any of the eigenvalues $A_n$, unless $A_n$ happens to be a moderate size integer, which is
not a typical situation. \\

Therefore, the \\

{\bf Question} :

\begin{quote}

To the extent that (6) cannot in general be rigorously confirmed, is the association in (8)
nevertheless {\it one-to-one} ?

\end{quote}

In other words, since in general the confirmation of (6) can only be of the form \\

(9) $~~~~~~ |~ \mbox{Prob} (A_n) ~-~ | < \alpha_n, \psi_{t - 0} > |^2 ~| ~\leq~ \epsilon_n,
                                       ~~~~ n \in {\bf N} $ \\

where $\epsilon_n > 0$ may depend on $\alpha_n$ and $\psi_{t - 0}$, can nevertheless the
eigenvalues (3) be determined uniquely ? \\

And to aggravate the situation, it should also be noted that (7) itself cannot in general be
perfectly confirmed by effective practical experiments. Indeed, it is one of the fundamental
assumptions of Quantum Mechanics, expressed explicitly in one of the axioms, that {\it no}
state of any quantum system can be directly observable. Consequently, since (7) is a
statement about such states, it can only be confirmed indirectly, namely, via results of
measurements on the corresponding observables ${\cal A}$. \\
And this bring us back to the above questions related to (6), (8) and (9). \\ \\

{\bf 3. Comments} \\

There exists a certain awareness about the {\it possibility} that the association in (8) may
fail to be one-to-one. For instance, in Davies [p. 66], we find :

\begin{quote}

"Although the choice of the operators is not unique, they must comply with the commutation
relations and bear the same functional relationship to each other as the corresponding
classical quantities (the correspondence principle). In practice the choice of $r$ and $-ihv$
for position and momentum is conventional. Most other operators follow from these."

\end{quote}

Two points can be noted in this respect. \\
First, the {\it correspondence principle} bring in additional constraints on the association
in (8), in case this association may indeed happen to fail to be one-to-one when considered in
terms of (6) and (7) alone. \\
Second, a further constraint to be noted is as follows. Assume that in (8) we have the
one-to-many association \\

(10) $~~~~~~ \begin{array}{l}
                       ~~~ ~\longmapsto~ A_1 \\
                       {\cal A} \\
                       ~~~ ~\longmapsto~ A_2 \\ \\ \\
                       ~~~ ~\longmapsto~ B_1 \\
                       {\cal B} \\
                       ~~~ ~\longmapsto~ B_2
              \end{array} $ \\

for certain given observables ${\cal A}$ and ${\cal B}$. Then we shall have to require that \\

(11) $~~~~~~ A_1,~ B_1 ~~\mbox{commute} ~~~\Longleftrightarrow~~~
                                            A_2,~ B_2 ~~\mbox{commute} $ \\

and in addition, also the relations, Gillespie [p. 67] \\

(12) $~~~~~~ \Delta A_i ~.~ \Delta B_i ~\geq~ | c_i |/ 2 ~>~ 0,~~~~ i ~=~ 1, 2 $ \\

if one of the pairs $A_i,~ B_i$, with $i = 1, 2$, does not commute, and instead, it satisfies
the relation \\

(13) $~~~~~~ A_i B_i ~-~ B_i A_i ~=~ c_i I $ \\

where $c_i \in {\bf C},~ c_i \neq 0$ and $I$ is the identity operator on $H$. We recall that
$\Delta A_i$ and $\Delta B_i$ denote, Gillespie [p. 55], the uncertainties in the operators
$A_i$ and $B_i$, respectively. \\ \\

{\bf 4. Reformulation of the Question} \\

In view of section 3, we can reformulate the question in section 2, as follows. \\
Assume that, according to the correspondence principle, we identify the set of quantum
observables with the set \\

(14) $~~~~~~ {\cal H} ( H ) $ \\

of Hermitian operators on the given Hilbert space $H$. In other words, every Hermitian
operator $A \in {\cal H} ( H )$ is {\it labelled} by a unique observable ${\cal A}$, and
through this labelling all observables will become the label of a certain Hermitian operator.
Thus we assume the existence of a one-to-one mapping \\

(15) $~~~~~~ A ~\longleftrightarrow~ {\cal A} $ \\

The the only condition required on the mapping in (15) is to satisfy all the axioms of Quantum
Mechanics as formulated, for instance, in Gillespie. \\
When it comes to the eigenvalues and eigenvectors of the respective operators $A$, that will
mean that (6) and (7) alone must hold. \\

And then we are led to the \\

{\bf Reformulated Question} :

\begin{quote}

Is it possible to {\it relabel} in (15) the Hermitian operators according to

\end{quote}

(16) $~~~~~~ A^\prime ~\longleftrightarrow~ {\cal A} $

\begin{quote}

in such a way that \\

~~*)~~~ the relations (6) and (7) still hold \\

~**)~~~ the correspondence principle, and in particular (11),

\hspace*{1cm} (12) and (13) also hold \\

***)~~ the operator $A^\prime$ in (15) need {\it not} be a unitary

\hspace*{1cm} transformation of $A$ in (15) ?

\end{quote}

{\bf Acknowledgment.} Thanks are due to my colleague Johan Swart for his comments on the first
two sections, which led to the clarifications presented in the last two sections.

\end{document}